\documentclass[12pt]{iopart}

\usepackage{bm}
\usepackage{dcolumn}
\usepackage{longtable}
\usepackage{multirow}
\usepackage{iopams}
\usepackage{graphicx}

\newcommand{\lambdabar}{\lambda\kern-.5em\raise.5ex\hbox{--}}

\newcommand{\Ninej}[9]{\left\{\matrix{
 {#1} & {\!\!\!#2} & {\!\!\!#3}\cr
 {#4} & {\!\!\!#5} & {\!\!\!#6}\cr
 {#7} & {\!\!\!#8} & {\!\!\!#9}\cr}\right\}}

\newcommand{\Cleb}[6]{C^{{\,#1}{\,#2}{\,#3}}
  _{{\,#4}{\,#5}{\,#6}} }

\newcommand{\qnfrac}{\frac}

\begin{document}

\title[Long-range states in excited ultracold $^{3}$He$^*$--$^{4}$He$^*$ dimers]
{Long-range states in excited ultracold $^{3}$He$^*$--$^{4}$He$^*$ dimers}

\author{D G Cocks$^1$, G Peach$^2$ and I B Whittingham$^1$}

\address{$^1$ College of Science, Technology and Engineering, James Cook 
University, Townsville 4811, Australia} 

\address{$^2$ Department of Physics and Astronomy, University College
London, London WC1E~6BT, UK}

\ead{daniel.cocks@jcu.edu.au}

\begin{abstract}
Long-range bound states of the excited heteronuclear 
${}^{3}$He$^{*}$--${}^{4}$He$^{*}$ system that dissociate to either 
${}^{3}$He(1s2s~${}^{3}$S$_{1}$) + ${}^{4}$He(1s2p~${}^{3}$P$_{j}$) or
${}^{3}$He(1s2p~${}^{3}$P$_{j}$) + ${}^{4}$He(1s2s~${}^{3}$S$_{1}$),
where $j=0, 1, 2$, are investigated using both single-channel and multichannel 
calculations in order to analyse the effects of Coriolis and non-adiabatic 
couplings. The multichannel calculations predict two groups of resonances above 
the lowest asymptotic energy. One of these groups dissociates to an atomic pair 
with the $2p$ excitation on the fermionic atom and the other dissociates to two 
asymptotes which correspond to the $2p$ excitation on either atom. Many of these 
resonances could be identified with levels in the single-channel calculation 
although the differences in energies were large.  The total parity was found to 
have a significant influence on the ability to make these identifications. No
purely bound states were found, although several resonances with line widths 
smaller than 1~MHz were obtained.

\end{abstract}


\submitto{\jpb}
\maketitle

\section{Introduction}

A widely used technique to study the dynamics of ultracold collisions in dilute 
quantum gases is photoassociation in which two interacting ultracold atoms are 
resonantly excited by a laser to bound states of the molecule formed during the 
collision.  Photoassociation in metastable rare gases is of particular interest 
since novel experimental strategies can be implemented.  In contrast to 
ground-state atomic species which have been trapped at ultracold temperatures, 
the large internal energy of the metastable atom can allow experimentalists to 
easily detect individual events with high resolution and hence potentially count 
each atom which has ionized or escaped from the trap \cite{Vassen12,Jeltes07}.

Photoassociation of two bosonic metastable ${}^{4}$He$^{*}$ atoms,
${}^{4}$He(1s2s~${}^{3}$S$_1$), to excited rovibrational bound states that 
dissociate to the ${}^{4}$He(1s2s~${}^{3}$S$_1$) + 
${}^{4}$He(1s2p~${}^{3}$P$_{j}$)
limits, where $j=0,1,2$, and of two fermionic metastable ${}^{3}$He$^{*}$ atoms,
${}^{3}$He(1s2s~${}^{3}$S$_1$), to states that dissociate to 
${}^{3}$He(1s2s~${}^{3}$S$_1$) + ${}^{3}$He(1s2p~${}^{3}$P$_{j}$), have recently 
been theoretically investigated by Cocks \textit{et al.} \cite{CWP10,CPW11}.
They use both single-channel and multichannel calculations based upon
input molecular potentials constructed from the short-range 
\textit{ab initio} ${}^{1,3,5}\Sigma^{+}_{g,u}$ and ${}^{1,3,5}\Pi_{g,u}$ 
potentials of Deguilhem \textit{et al.}\cite{DLGD09} matched onto long-range 
retarded resonance dipole and dispersion potentials. 

The multichannel calculations permitted criteria to be established 
for the assignment of the theoretical levels to experimental observations in the 
case of the bosonic system and, in the absence of any experimental observations 
of bound states in the fermionic system, predictions as to which of the 
calculated bound states may be experimentally observable. 

The heteronuclear ${}^{3}$He$^{*}$--${}^{4}$He$^{*}$ system has been relatively 
unexplored. The spin polarized mixture with ${}^{3}$He(1s2s~${}^{3}$S$_1$) in
the state $|f,m_{f}\rangle = |3/2,+3/2 \rangle$ and  
${}^{4}$He(1s2s~${}^{3}$S$_1$) in the state $|j,m_{j}\rangle =|1,+1 \rangle $
has been simultaneously magneto-optically trapped \cite{SMHV04} and 
${}^{4}$He$^{*}$ used to sympathetically cool ${}^{3}$He$^{*}$ to the 
quantum degenerate regime \cite{MJTHV06}. Goosen \textit{et al.} \cite{GTVK10} 
have undertaken a theoretical investigation of Feshbach resonances in 
homonuclear and heteronuclear mixtures of  ${}^{3}$He$^{*}$ and ${}^{4}$He$^{*}$,
predicting a broad resonance in the heteronuclear system. Recently,
Borbely \textit{et al.} \cite{BRKV12} have predicted that, as 
${}^{3}$He$^{*}$--${}^{4}$He$^{*}$ mixtures prepared in their lowest spin 
channel are stable against Penning ionization, they provide the
ideal starting point for future experiments such as preparing an 
ultracold mixture in an optical dipole trap in order to study this
Feshbach resonance. 

We report here a theoretical investigation of the long-range bound states
of the excited heteronuclear ${}^{3}$He$^{*}$--${}^{4}$He$^{*}$ system that 
dissociate to either 
${}^{3}$He(1s2s~${}^{3}$S$_{1}$) + ${}^{4}$He(1s2p~${}^{3}$P$_{j}$) or
${}^{3}$He(1s2p~${}^{3}$P$_{j}$) + ${}^{4}$He(1s2s~${}^{3}$S$_{1}$),
where $j=0, 1, 2$. 
The structure of these states is expected to differ substantially from those
found for the homonuclear systems as the large number of asymptotes available
in the heteronuclear configuration means many of the bound states of a given
channel will be embedded in a continuum of states of other channels and become 
resonances associated with predissociation.
These states are relevant to any future studies of 
photoassociation in such mixtures and are investigated using both single-channel 
and multichannel calculations in order to analyse the effects of Coriolis and 
non-adiabatic couplings.

Atomic units are used, with lengths in Bohr radii $a_{0} = 0.0529177209$ nm 
and energies in Hartree $E_{h}=\alpha ^{2}m_{e}c^{2}=27.211384$ eV.

\section{Theory}
\subsection{Multichannel equations}

The formalism for the excited heteronuclear 
\mbox{${}^{3}$He$^{*}$--${}^{4}$He$^{*}$} system requires some modification of 
that presented by Cocks \textit{et al.} \cite{CPW11} for the excited homonuclear 
\mbox{${}^{3}$He$^{*}$--${}^{3}$He$^{*}$} system. In particular, there are fewer
symmetries in the heteronuclear system that can be taken advantage of and we 
must be sure to include all coupled states, including those states possessing 
asymptotic forms with the excitation on either the fermionic or bosonic atom.

The total Hamiltonian for a system of two interacting atoms $i=1,2$ with 
reduced mass $\mu $, interatomic separation $R$ and relative angular momentum 
$\hat{\bm{l}}$, where both atoms possess fine structure and one hyperfine 
structure is 
\begin{equation}
\label{cpw1}
\hat{H} = \hat{T}_K + \hat{H}_\mathrm{rot} + \hat{H}_\mathrm{el} +
\hat{H}_\mathrm{fs} + \hat{H}_\mathrm{hfs}
\end{equation}
where $\hat{T}_K$ is the kinetic energy operator
\begin{equation}
\label{cpw2}
\hat{T}_K=-\frac{\hbar ^{2}}{2\mu R^{2}}\frac{\partial}{\partial R}
\left( R^{2}\frac{\partial}{\partial R}\right)\,
\end{equation}
and $\hat{H}_{\mathrm{rot}}$ the rotational operator
\begin{equation}
\label{cpw3}
\hat{H}_{\mathrm{rot}} = \frac{\hat{l}^{2}}{2 \mu R^{2}}. 
\end{equation}
The total electronic Hamiltonian is 
\begin{equation}
\label{cpw4}
\hat{H}_{\mathrm{el}}=\hat{H}_{1}+\hat{H}_{2}+\hat{H}_{12},
\end{equation}
where $\hat{H}_{i}$ is the unperturbed Hamiltonian of atom $i$ and 
$\hat{H}_{12}$ is the electrostatic interaction between the atoms. The terms 
$\hat{H}_{\mathrm{fs}}$ and $\hat{H}_\mathrm{hfs}$  in equation (\ref{cpw1}) 
describe the fine structure and hyperfine structure respectively of the atoms.  

The multichannel equations describing the interacting atoms are obtained by 
writing the eigenvector $|\Psi \rangle $ of the total system, which satisfies
\begin{equation}
\label{cpw5}
\hat{H} |\Psi \rangle = E |\Psi \rangle ,
\end{equation}
in terms of an expansion
\begin{equation}
\label{cpw6}
|\Psi \rangle = \sum_a \frac{1}{R} G_a(R) |a \rangle  ,
\end{equation}
where $G_a(r)$ are vibrational wave functions and the molecular basis is  
$|a\rangle = |\Phi_a (R,q) \rangle$, where $q$ denotes the interatomic polar 
coordinates $(\theta ,\varphi)$ and electronic coordinates $\{\bm{r}_{i}\}$. The 
state label, $a$, denotes the set of approximate quantum numbers describing the 
electronic-rotational states of the molecule.  We make the Born-Oppenheimer (BO)
approximation that the basis states $|a\rangle$ depend only parametrically on 
$R$ so that $\langle a^\prime| \hat{T}_K |a\rangle = 0$. In this approximation 
\cite{Gao96}, the nuclear separation only enters the matrix elements through the 
molecular potentials of the electronic Hamiltonian $\hat{H}_\mathrm{el}$ and the 
rotational Hamiltonian $\hat{H}_\mathrm{rot}$.  This implies the 
hyperfine-structure is $R$-independent and that the bulk of the $R$-dependence 
of $|\Psi\rangle$ is contained in the vibrational factors $G_a(R)$. We have 
previously used this approximation to achieve very good agreement between our 
calculations and the many experimental observations for the homonuclear 
\mbox{${}^4$He${}^*$--${}^4$He${}^*$} system \cite{CWP10}. We note that it is 
common to refer to BO states as adiabatic states in the absence of hyperfine 
structure and rotational couplings. In our case, the states $|a\rangle$ are a 
unitary transformation from the set of BO states and we 
later refer to adiabatic states of the complete Hamiltonian neglecting the 
kinetic energy term, $\hat{T}_K$.
Forming the scalar product $\langle a^{\prime}|\hat{H}|\Psi \rangle $ yields the 
set of multichannel equations
\begin{equation}
\label{cpw7}
\sum_a \left[-\frac{\hbar^{2}}{2\mu }\frac{d^{2}}{dR^{2}}
\delta_{a^{\prime},a} + V_{a^\prime a}(R) - E 
\delta_{a^{\prime},a}\right] G_{a}(R) = 0\,,
\end{equation}
where
\begin{equation}
\label{cpw9}
V_{a^{\prime} a}(R) = \langle a^{\prime}| \left[\hat{H}_{\mathrm{rot}} + 
\hat{H}_{\mathrm{el}} + \hat{H}_{\mathrm{fs}} + \hat{H}_{\mathrm{hfs}}\right] 
|a \rangle .
\end{equation}

\subsection{Basis states and matrix elements}

The excited heteronuclear ${}^{3}$He$^{*}$--${}^{4}$He$^{*}$ system can be in 
two possible arrangements: ${}^{3}$He(1s2s~${}^{3}$S$_{1}$) + 
${}^{4}$He(1s2p~${}^{3}$P$_{j}$) and ${}^{3}$He(1s2p~${}^{3}$P$_{j}$) + 
${}^{4}$He(1s2s~${}^{3}$S$_{1}$).

For notational convenience we shall assume both nuclei have angular momentum 
$\hat{\bm{i}}_{i}$ and set the appropriate nuclear angular momentum to zero at 
the end of the formalism.  If the two colliding atoms have orbital 
$\hat{\bm{L}}_{i}$ and spin $\hat{\bm{S}}_{i}$ angular momenta, the body-fixed 
eigenstates of total parity $\hat{P}_{T}$ for the two arrangements in the 
coupling scheme
\begin{equation}
\label{cpw11}
\hat{\bm{j}}_{i}=\hat{\bm{L}}_{i}+\hat{\bm{S}}_{i},  \quad 
\hat{\bm{f}}_{i}=\hat{\bm{j}}_{i}+\hat{\bm{i}}_{i},  \quad
\hat{\bm{f}}= \hat{\bm{f}}_{1}+\hat{\bm{f}}_{2},     \quad
\hat{\bm{T}}=\hat{\bm{f}}+\hat{\bm{l}}
\end{equation}
are (see \ref{app:states} for details)
\begin{eqnarray}
\label{cpw12}
|a \rangle_{12} & \equiv |(\alpha_1)_A,(\alpha_2)_B,f,\phi,T,m_T,P_T\rangle 
\nonumber \\
				& \equiv |(\gamma_{1}j_{1}i_{A}f_{1A})_{A},
				(\gamma_{2}j_{2}i_{B}f_{2B})_{B}, f, \phi, T, m_{T}; 
				P_{T}\rangle
\end{eqnarray}
and
\begin{eqnarray}
\label{cpw12a}
|a \rangle_{21} & \equiv |(\alpha_2)_A,(\alpha_1)_B,f,\phi,T,m_T,P_T\rangle 
\nonumber \\
				& \equiv |(\gamma_{2}j_{2}i_{A}f_{2A})_{A},
				(\gamma_{1}j_{1}i_{B}f_{2B})_{B}, f, \phi, T, m_{T}; 
				P_{T}\rangle
\end{eqnarray}
where $\gamma_{i}\equiv \{\bar{\gamma}_{i},L_{i},S_{i}\}$, $\bar{\gamma}_{i}$ 
representing any other relevant quantum numbers, and  $\phi \equiv |\Omega_{f}| 
= |\Omega_{T}|$. The nuclei of $^{3}$He and $^{4}$He are labelled $A$ and $B$ 
respectively and we define the intermolecular axis to be $\bm{R} = \bm{r}_B - 
\bm{r}_A$. The sets of quantum numbers $(\gamma_{1},j_{1})$ and 
$(\gamma_{2},j_{2})$ describe the 1s2s~${}^{3}$S$_{1}$ and 1s2p~${}^{3}$P$_{j}$ 
states respectively.  The projections of an angular momentum $\hat{\bm{J}}$ onto 
the space-fixed axis $Oz$ and inter-molecular axis $OZ$ with orientation 
$(\theta,\varphi)$ relative to the space-fixed frame are denoted $m_{J}$ and 
$\Omega_{J}$ respectively.  

The alternative body-fixed states
\begin{equation}
\label{cpw18aa}
|(\gamma_{1})_{A}, (\gamma_{2})_{B}, L S \Omega_{L} \Omega_{S} \rangle 
\end{equation}
and
\begin{equation}
\label{cpw18b}
|(\gamma_{2})_{A}, (\gamma_{1})_{B}, L S \Omega_{L} \Omega_{S} \rangle 
\end{equation}
arising from the couplings $\hat{\bm{L}}=\hat{\bm{L}}_{1}+\hat{\bm{L}}_{2}$ 
and  $\hat{\bm{S}}=\hat{\bm{S}}_{1}+\hat{\bm{S}}_{2}$ are required in the 
evaluation of the matrix elements of $\hat{H}_{\mathrm{el}}$. The relationship 
between the two bases (\ref{cpw12}) and (\ref{cpw18aa}) is given by (see 
\ref{app:states})
\begin{eqnarray}
	\label{cpw19a}
	\fl
	|(\alpha_1)_A,(\alpha_2)_B,f,\phi,T,m_T,P_T\rangle = \nonumber \\
	|(\alpha_1)_A,(\alpha_2)_B,f,\Omega_f{=}\phi,T,m_T\rangle \nonumber \\
	{} + P_T (-1)^{f-T+1} 
	|(\alpha_1)_A,(\alpha_2)_B,f,\Omega_f{=}-\phi,T,m_T\rangle
\end{eqnarray}
and
\begin{eqnarray}
\label{cpw19}
\fl
|(\alpha_1)_{A}, (\alpha_2)_{B}, f, \Omega_f, T, m_{T} \rangle  \nonumber  
\\
= |T, m_{T}, \Omega_f \rangle \sum_{i \Omega_{i} j} 
\sum_{LS\Omega_{L}\Omega_{S}} \tilde{F}^{12}_{AB}
|(\gamma_{1})_{A}, (\gamma_{2})_{B}, L S \Omega_{L} \Omega_{S} \rangle |i_A, 
i_B, i \Omega_{i} \rangle .
\end{eqnarray}
The coupling coefficients $\tilde{F}^{12}_{AB}$ are defined in \ref{app:states} 
and include the quantum numbers 
$(L_1,L_2,L,S_1,S_2,S,j_{1},j_{2},j,i_A,i_B,i,f_{1A},f_{2B},f,\Omega_L,\Omega_S,\Omega_j,\Omega_i,\Omega_f)$ 
implicitly. The rotational states are
\begin{equation}
\label{cpw21}
|T,m_{T}, \Omega_f \rangle = \sqrt{ \frac{2T+1}{4 \pi }} 
D^{T\,*}_{m_{T},\Omega_f}(\varphi , \theta , 0),
\end{equation}
where $D^{T\,*}_{m_{T},\Omega_f}(\varphi ,\theta ,0)$ is the Wigner rotation 
matrix.  The analogous relationship between the states (\ref{cpw12a}) and 
(\ref{cpw18b}) follows from (\ref{cpw19}) by interchanging $1$ and $2$ but 
leaving $A$ and $B$ fixed.

The multichannel equations (\ref{cpw7}) require the matrix elements of 
$\hat{H}_{\mathrm{rot}}$, $\hat{H}_{\mathrm{el}}$, $\hat{H}_{\mathrm{fs}}$ and
$\hat{H}_{\mathrm{hfs}}$ in the basis $\{|a\rangle_{12},|a\rangle_{21}\}$.  The 
non-zero rotation terms are \begin{eqnarray}
\label{cpw24}
\fl
{}_{12}\langle a^{\prime}|\hat{l^{2}}|a \rangle_{12}  &=
{}_{21}\langle a^{\prime}|\hat{l^{2}}|a \rangle_{21} \nonumber  \\
&= \hbar^{2} \delta_{\rho , \rho^{\prime}}
\left\{\left[T(T+1)+f(f+1)-2\phi^{2}\right] \delta_{\phi^{\prime},\phi} 
-C_{\phi^{\prime}\phi}\right \}
\end{eqnarray} 
where
\begin{eqnarray}
\label{cpw24a}
C_{\phi^{\prime}\phi} = \left \{ \begin{array} {ll}
		P_{T}(-1)^{f-T-1}K^{-}_{Tf\phi} & \mathrm{for\;} \phi = 
		\phi^{\prime}=\qnfrac{1}{2} \\
K^{-}_{Tf\phi} \;\delta_{\phi^{\prime},\phi-1} +
K^{+}_{Tf\phi}\; \delta_{\phi^{\prime},\phi+1} & \mathrm{otherwise}
\end{array}
\right..
\end{eqnarray}
The Coriolis coupling terms are 
\begin{equation}
\label{cpw25}
\fl
K^{\pm}_{Tf\phi}  =  
\left[T(T+1) - \phi(\phi \pm 1)\right]^{\qnfrac{1}{2}}
\left[ f(f+1) - \phi (\phi \pm 1)\right]^{\qnfrac{1}{2}},
\end{equation}
and $\rho$ denotes the set of quantum numbers
$\{(\alpha_i)_A,(\alpha_j)_B,f,T, m_{T}, P_{T}\}$.

The electronic matrix elements can be expressed in terms of the BO molecular 
potentials ${}^{2S+1}\Lambda^{\sigma}_{w}(R)$, where $\Lambda 
\equiv|\Omega_{L}|$, which are identical to that of the homonuclear 
${}^{3}$He--${}^{3}$He or ${}^{4}$He--${}^{4}$He systems. The Born-Oppenheimer 
potentials are eigenvalues of $\hat{H}_\mathrm{el}$ for the symmetrized states:
\begin{eqnarray}
\label{cpw18}
\fl
|\gamma_{1}, \gamma_{2},L S \Omega_{L} \Omega_{S};w \rangle =
N_{w}\left[|(\gamma_{1})_{A}, (\gamma_{2})_{B}, L S \Omega_{L} \Omega_{S} \rangle 
\right.
\nonumber  \\
+(-1)^{w}\varepsilon_{LS} \left.
|(\gamma_{2})_{A}, (\gamma_{1})_{B}, L S \Omega_{L} \Omega_{S} \rangle \right]
\end{eqnarray}
where $N_{w}=1/\sqrt{2(1+\delta_{\gamma_{1},\gamma_{2}})}$, $w=0(1)$ for
\textit{gerade} (\textit{ungerade}) symmetry and
\begin{equation}
\label{cpw18a}
\varepsilon_{LS}=(-1)^{L_{1}+L_{2}-L +S_{1}+S_{2}-S+N_{1}N_{2}}P_{1}P_{2}.
\end{equation}
Here $P_{i}=(-1)^{L_{i}}$ is the parity of the atomic state 
$|L_{i}m_{L_{i}}\rangle $ and $N_{i}$ is the number of electrons on atom $i$. 
Using
\begin{equation}
\label{cpw26}
\fl 
\hat{H}_{\mathrm{el}}
|\gamma_{1}, \gamma_{2},L S \Omega_{L} \Omega_{S};w \rangle =
\left[{}^{2S+1}\Lambda^{\sigma}_{w}(R)+ E_{\Lambda S}^{\infty}\right]
|\gamma_{1}, \gamma_{2},L S \Omega_{L} \Omega_{S};w \rangle 
\end{equation}
where $E_{\Lambda S}^{\infty}$ is the asymptotic energy of the state, the matrix 
elements of $\hat{H}_{\mathrm{el}}$ are (see \ref{app:states})
\begin{eqnarray}
\label{cpw27}
\fl
{}_{12}\langle a^{\prime}|\hat{H}_{\mathrm{el}}|a \rangle_{12}  =  
\delta_{\eta^{\prime},\eta} \sum_{i,\Omega_{i}} \sum_{j^{\prime},j}
\sum_{LS\Omega_{L}\Omega_{S}} \tilde{F}^{12}_{AB} \tilde{F}^{1^\prime 
2^\prime}_{AB} \nonumber  \\
\times\frac{1}{2} [{}^{2S+1}\Lambda_{g}^{+}(R)+{}^{2S+1}\Lambda_{u}^{+}(R)+
2E_{\Lambda S}^{\infty}]
\end{eqnarray}
where $1^\prime 2^\prime$ corresponds to the quantum numbers of $a^\prime$.  The 
matrix elements for ${}_{21}\langle a^\prime|\hat{H}_\mathrm{el}|a\rangle_{21}$ 
are obtained by interchanging $1\leftrightarrow 2$ in (\ref{cpw27}) and the 
cross terms are:
\begin{eqnarray}
\label{cpw27a}
\fl
{}_{21}\langle a^{\prime}|\hat{H}_{\mathrm{el}}|a \rangle_{12}  =  
\delta_{\eta^{\prime},\eta} \sum_{i,\Omega_{i}} \sum_{j^{\prime},j}
\sum_{LS\Omega_{L}\Omega_{S}} \varepsilon_{LS} \tilde{F}^{2^\prime 1^\prime}_{AB}
\tilde{F}^{12}_{AB} \nonumber  \\
\times\frac{1}{2} [{}^{2S+1}\Lambda_{g}^{+}(R)-{}^{2S+1}\Lambda_{u}^{+}(R)]
\end{eqnarray}
where $\eta$ denotes the set of quantum numbers 
$\{\gamma_{1},\gamma_{2},\phi,T,m_{T}, P_{T}\}$. The summations over 
$\{i,L\}$ disappear after the zero angular momentum of the ${}^{4}$He nucleus 
and the 1s2s~${}^{3}$S$_{1}$ state quantum number
$L_{1}=0$ are assigned.

We assume that the fine and hyperfine structure of the individual atoms is 
unaffected by the formation of the dimer, so that \begin{equation}
\label{cpw28}
\fl
{}_{12}\langle a^\prime | \hat{H}_\mathrm{fs} + \hat{H}_\mathrm{hfs} | a 
\rangle_{12} = \delta_{a^\prime, a} \Delta E^\mathrm{fs}_{\gamma_2 j_2}  + 
\delta_{\sigma^\prime , \sigma} \langle (\alpha_{1}^\prime)_A | 
\hat{H}_\mathrm{hfs} | (\alpha_{1})_A \rangle \delta_{(\alpha_{2}^\prime)_B 
,(\alpha_{2})_B} \end{equation}
and
\begin{equation}
\label{cpw28a}
\fl
{}_{21}\langle a^\prime | \hat{H}_\mathrm{fs} + \hat{H}_\mathrm{hfs} | a 
\rangle_{21} = \delta_{a^\prime, a} \Delta E^\mathrm{fs}_{\gamma_1 j_1}  + 
\delta_{\sigma^\prime , \sigma} \langle (\alpha_{2}^\prime)_A | 
\hat{H}_\mathrm{hfs} | (\alpha_{2})_A \rangle \delta_{(\alpha_{1}^\prime)_B 
,(\alpha_{1})_B}, \end{equation}
where $\sigma$ denotes the set of quantum numbers $\{f, \phi, T, m_T, P_{T}\}$. 
The fine structure splittings $\Delta E^\mathrm{fs}_{\gamma_2 j_2}$ for 
the 2p ${}^{3}$P$_{j_{2}}$ states and the hyperfine structure matrix elements 
$\langle \alpha^\prime| \hat{H}_\mathrm{hfs} | \alpha \rangle $ for the 
${}^{3}$He nucleus are taken from Wu and Drake \cite{Wu07} and the
hyperfine splitting of the 1s2s~${}^{3}$S$_{1}$ ${}^{3}$He level from Zhao 
\textit{et al.} \cite{Zhao91}.
For the 1s2p configuration of ${}^{3}$He there are seven relevant singlet and 
triplet states
\mbox{$|\alpha \rangle = |\bar{\gamma},L=1,S,j,i=\qnfrac{1}{2},f \rangle \equiv 
|S,j,f \rangle$}.
As $\hat{H}_\mathrm{hfs}$ does not couple states with different $f$ values, these seven
states form three sets $\{|0,1,\qnfrac{1}{2} \rangle, |1,0,\qnfrac{1}{2} \rangle , 
|1,1,\qnfrac{1}{2} \rangle\}$,
$\{|0,1,\qnfrac{3}{2} \rangle, |1,1,\qnfrac{3}{2} \rangle , |1,2,\qnfrac{3}{2} \rangle \}$ and 
$\{|2,2,\qnfrac{5}{2} \rangle \}$. The $7 \times 7$ matrix 
$A_{\alpha^{\prime}\alpha} \equiv \langle 
S^{\prime},j^{\prime},f|\hat{H}_\mathrm{hfs}|S,j,f \rangle $ is hence block
diagonal with $3 \times 3$ blocks for $f=\qnfrac{1}{2}$ and $f=\qnfrac{3}{2}$ and a single 
element
for $f=\qnfrac{5}{2}$. Diagonalization of the subblocks, which include the singlet states, 
is necessary to obtain accurate asymptotic energies, as the coupling between 
singlet and triplet states shifts the energies noticeably.
However, we do not want to include the ``dressed'' singlet states (i.e. the 
states after diagonalization that are close to the uncoupled singlet state 
energies) in our multichannel basis as they are well separated in energy from the 
triplet states and make a negligible contribution to the scattering calculation.
Fortunately, the eigenstates $|\beta, f \rangle 
=\sum_{S,j}U^{f,\beta}_{S,j}|S,j,f \rangle $
resulting from the diagonalization of  $A_{\alpha^{\prime}\alpha}$ can be 
labelled in terms of approximate quantum numbers $(\tilde{S},\tilde{j}, f)$
associated with the state $|\tilde{S},\tilde{j},f \rangle$ which has the 
largest projection onto $|\beta ,f \rangle $, that is
$|\beta, f \rangle \equiv |\tilde{S}, \tilde{j}, f \rangle \equiv 
|\tilde{\alpha}\rangle $. Hence, we choose to neglect those eigenstates with 
predominantly singlet character ($\tilde{S}=0$), which we justify by noting that 
the contribution of the original
singlet states $|S=0,j,f \rangle $ to the $|\tilde{S}=1,\tilde{j},f \rangle $
states is negligible (amplitude $< 10^{-7}$). The final states are written 
explicitly as:
\begin{eqnarray}
	\fl
	|\tilde{S}_i = 1,\tilde{j}_i = f_i \pm \qnfrac{1}{2}, f_i \rangle &= 
	\cos(\theta_{f_i})|\gamma_i, j_i = f_i \pm \qnfrac{1}{2}, f_i\rangle \nonumber 
	\\
	& {} \mp \sin(\theta_{f_i}) |\gamma_i, j_i=f_i \mp \qnfrac{1}{2},f_i\rangle,
\end{eqnarray}
for $f_i=\qnfrac{1}{2}$ and $f_i=\qnfrac{3}{2}$, where $\theta_{f_i}$ is determined from the 
eigenstates, and
\begin{equation}
	|\tilde{S}_i = 1,\tilde{j}_i = 2, f_i = 5/2 \rangle = 
	|\gamma_i,j_i=2,f_i=5/2\rangle
\end{equation}
The hyperfine energies given in table \ref{tab:levels} define a diagonal matrix 
in this new basis $|\tilde{\alpha}_i \rangle$ and are transformed into the 
$|\alpha_i\rangle$ basis to be used in (\ref{cpw28}) for our numerical 
calculations.

\begin{table}
\caption{\label{tab:levels} Fine and hyperfine structure of the
${}^{3}$He--${}^{4}$He system. The hyperfine energies $E^\mathrm{hfs}$,
in MHz, of ${}^{3}$He for the 2s ${}^{3}$S states are given relative to the
$(j,f)=(1,3/2)$ state and those for the 2p ${}^{3}$P states relative to 
the $(j,f)=(2,5/2)$ state. The fine structure energies $E^\mathrm{fs}$,
in MHz, of ${}^{4}$He for the 2p ${}^{3}$P states are given relative to the 
$j=2$ state. The fine structure energies $\tilde{E}^\mathrm{fs}$ have been 
shifted upwards so that the centres of gravity of the 2s ${}^{3}$S
and 2p ${}^{3}$P manifolds are the same for ${}^{3}$He and ${}^{4}$He.}
\begin{indented}
\lineup
\item[]\begin{tabular}{lllllll}
\br
& \centre{3}{${}^{3}$He} & \centre{3}{${}^{4}$He}  \\
& \crule{3} & \crule{3}  \\
Level & $j$ & $f$ & $E^\mathrm{hfs}$ & $j$ & $E^\mathrm{fs}$ & 
$\tilde{E}^\mathrm{fs}$ \\
2p ${}^{3}$P & 0 & $\qnfrac{1}{2}$ & 34385.941 & 0 & 31908.83798 & 34061.91194  \\
             & 2 & $\qnfrac{3}{2}$ & \06961.104 & & & \\
             & 1 & $\qnfrac{1}{2}$ & \06293.071 & 1 & \02292.16354 & \04445.23575 \\
             & 1 & $\qnfrac{3}{2}$ & \01780.880 & & & \\
             & 2 & $\qnfrac{5}{2}$ & \00        & 2 & \00 & \02153.07216 \\
&&&&&& \\
2s ${}^{3}$S & 1 & $\qnfrac{1}{2}$ & \06739.701177 & & & \\
             & 1 & $\qnfrac{3}{2}$ & \00 & 1 &  \00 &  \02246.5671  \\
\br
\end{tabular}
\end{indented}
\end{table}

The total matrix element $V_{a^{\prime}a}(R)$ is therefore diagonal in $T$ and 
$m_T$ and its non-zero values are furthermore independent of $m_{T}$.

\subsection{Single-channel approximation}

To make a single-channel approximation, we first neglect the Coriolis couplings
in (\ref{cpw24}) to obtain sets of channels for each value of $\phi$. We then 
find the adiabatic states within each of these sets by diagonalizing at each 
value of $R$, the matrix
\begin{equation}
\label{cpw34}
V_{a\prime a}^{\phi} = 
\langle a^{\prime} | \hat{H}_{\mathrm{el}}|a \rangle + 
\langle a^{\prime}|(\hat{H}_{\mathrm{fs}}+\hat{H}_{\mathrm{hfs}})|a \rangle
+\frac{\langle a^{\prime} |\hat{l}^{2} |a \rangle_{\phi}} {2 \mu R^{2}} \,,
\end{equation}
where $|a \rangle \equiv \{|a \rangle_{12}, |a \rangle_{21}\}$ and $\langle 
a^{\prime} |\hat{l}^{2} |a \rangle _{\phi}$ is the part of (\ref{cpw24}) 
with $C_{\phi^\prime \phi}$ (i.e. the Coriolis couplings) neglected. The 
corresponding $R$-dependent eigenvectors are
\begin{equation}
\label{cpw35}
|n \rangle = \sum_{a} C_{an} (R) |a \rangle ,
\end{equation}
where $n=\{\phi ,T, k \}$, which includes an index $k = 0,1,2,...$ (assigned in 
order of increasing energy at large $R$) to distinguish the different 
eigenvectors of the subspace $\{\phi,T\}$, and the adiabatic potential is given 
by \begin{equation}
\label{cpw34a}
V^\mathrm{adi}_n(R) = 
\sum_{a^{\prime}a} C^{-1}_{a^{\prime}n}V^{\phi}_{a^{\prime}a}C_{an}.
\end{equation}
Note that the eigenvectors are degenerate in $m_T$ and $P_T$, which have been 
omitted.  The radial eigenvalue equation for the rovibrational eigenstates 
$|\psi_{n,v} \rangle = R^{-1}G_{n,v}(R)|n \rangle$ is then
\begin{equation}
\label{cpw37}
\left[ -\frac{\hbar^2}{2\mu }\frac{\rmd^2}{\rmd R^2} + V^\mathrm{adi}_{n}(R) 
- E_{n,v} \right] G_{n,v}(R) = 0\,.
\end{equation}
We define the couplings that have been neglected in (\ref{cpw37}), which arise 
from $\hat{T}_K$ and are hence proportional to the derivatives $dC_{an}/dR$ and 
$d^2C_{an}/dR^2$, as non-adiabatic couplings. These couplings 
are important only when both the diagonalization of the potential varies quickly 
with $R$ and the energy difference between two adiabatic potentials is small.

We note here that there exist no special symmetries that cause couplings to 
disappear at particular values of $R$. Hence, none of the adiabatic potentials 
within one particular set cross one another. However, it will be seen in the 
multichannel calculations that the non-adiabatic terms can cause diabatic 
transitions between these potentials, leading to effective true crossings 
between these adiabatic potentials.

\subsection{Input potentials}

The required Born-Oppenheimer potentials ${}^{1,3,5}\Sigma^{+}_{g,u}$ and 
${}^{1,3,5}\Pi_{g,u}$ were constructed as in Cocks \textit{et al.}
\cite{CWP10,CPW11} by matching the \textit{ab initio} short-range 
potentials of Deguilhem \textit{et al.}\cite{DLGD09} onto the long-range 
dipole-dipole plus dispersion potentials 
\begin{eqnarray}
\label{cpw38}
V_{\Lambda }^{\mathrm{long}}(R)= -f_{3\Lambda}(R/\lambdabar)
C_{3\Lambda}/R^{3} - C_{6\Lambda}/R^{6}
\nonumber  \\ - C_{8\Lambda}^\pm / R^{8} - 
C_{9\Lambda} /R^{9} - C_{10\Lambda}/R^{10},
\end{eqnarray}
where $f_{3\Lambda}$ is an $R$- and $\Lambda$-dependent retardation 
correction~\cite{Meath68}, $\lambdabar=\lambda/(2\pi) = 3258.12 a_{0}$ 
where $\lambda$ is the wavelength for the 2s$\,{}^3$S--2p$\,{}^3$P 
transition and the parameters $C_{n \Lambda}$ were taken from  Zhang 
\textit{et al.}\cite{Zhang06}. Again, motivated by our study of the 
${}^{4}$He$^{*}$--${}^{4}$He$^{*}$ system \cite{CWP10}, we apply a 1\% increase 
to the slope of the ${}^{5}\Sigma^{+}_{g,u}$ and ${}^{5}\Pi_{g,u}$ potentials 
near their inner classical turning point. In our previous calculations, this 1\%
increase produced excellent agreement between many of the theoretical and 
experimental results and brought most of the theoretical predictions to well 
within the $20$~MHz uncertainty of the experimental measurements.

The coefficients $C_{3\Lambda}$ are of opposite sign for the $u$ and $g$ 
potentials. Consequently the dipole-dipole contribution is cancelled in the 
matrix elements ${}_{12}\langle a^{\prime}|\hat{H}_{\mathrm{el}}|a \rangle_{12}$ 
and  ${}_{21}\langle a^{\prime}|\hat{H}_{\mathrm{el}}|a \rangle_{21}$ as 
expected and only contributes to the off-diagonal elements  ${}_{21}\langle 
a^{\prime}|\hat{H}_{\mathrm{el}}|a \rangle_{12}$.  

\section{Results}

\subsection{Method}


Our numerical calculations follow closely those described in \cite{CWP10,CPW11} 
and we briefly outline the procedure here. The numerical solution of the coupled 
multichannel equations (\ref{cpw7}) and each single-channel equation 
(\ref{cpw37}) for a single energy $E$ is performed using the renormalized 
Numerov method on a grid of points consisting of connected regions with fixed 
step sizes. To obtain the single-channel bound states, we select only those 
potentials which have a minimum at long-range ($R>100\;a_0$) and determine the 
bound state eigenenergies by counting the number of nodes in the wave function 
as a function of energy for energies less than the asymptotic energy of the
single-channel, $E^\infty_n = V^\mathrm{adi}_n(R\rightarrow \infty)$.  The bound
state energies then correspond to a change in the number of nodes (or
equivalently a node at $R\rightarrow \infty$).

In the multichannel calculation, true bound states are unlikely to occur for 
energies above at least one of the channel asymptotes as couplings to open 
channels provide paths for dissociation. Hence, we extend our search to complex 
energy space $\mathcal{E} = E - i\Gamma$, where the value of $\Gamma$ at a valid 
resonance designates the line width of the resonance. We identify these 
resonances by performing inward and outward integration of the equations 
(\ref{cpw7}) and use the inverse of the matching condition as the integrand of a 
contour integration.  See \ref{app:MC} for details.

\begin{table}
\caption{\label{tab:adi} Single-channel rovibrational binding energies, 
in units of MHz, of long-range states in the
\mbox{${}^{3}$He(2 $^{3}$S$_{1}$)+${}^{4}$He(2 $^{3}$P$_{j}$)} and
${}^{3}$He(2 ${}^{3}$P$_{j}$) + ${}^{4}$He(2 ${}^{3}$S$_{1}$) 
systems. Energies given are relative to the specified asymptotic energy 
$E_{n}^{\infty}$ and the index of the adiabatic potential given by $k$.}
\begin{indented}
\lineup
\item[]\begin{tabular}{lllllllll}
\br
$\phi$ & $k$ & $E_{n}^{\infty}$ & $v/T$  & $\qnfrac{1}{2}$  & $\qnfrac{3}{2}$  & $\qnfrac{5}{2}$  & $\qnfrac{7}{2}$  &  $\qnfrac{9}{2}$ \\
\mr

$\qnfrac{1}{2}$ & 13 & 8539.6 & 0 & 1311.1  & 1274.2 & 1213.0  & 1128.2 & 1020.6 \\
	&    &   & 1 & \0517.0 & \0485.5 & \0434.3 & \0364.8 & \0279.7 \\
	&    &  & 2 & \0130.8 & \0111.4 & \0\080.4 & \0\040.4 & \\
	&    &  & 3 & \0\0\03.3 & & & & \\
	& 14 & 8539.6 & 0 & 1035.8  & 1002.0 & \0946.2 & \0868.9 & \0770.7 \\
	&    &  & 1 & \0325.4 & \0297.7 & \0252.7 & \0192.0 & \0117.8 \\
&    &  & 2 & \0\0\03.7 & & & & \\
\\
	& 15 & 8892.8 & 0 & \0516.6 & \0478.2 & \0415.3 & \0330.1 & \0226.6 \\
	&    &   & 1 & \0202.2 & \0185.4 & \0158.2 & \0121.5 & \0\075.8 \\
	&    &   & 2 & \0\051.2 & \0\039.8 & \0\022.1 &  &  \\
\\
	& 16 & 8892.8 & 0 & \0133.1 & \0103.7 & \0\056.0 &  &  \\
\\
	& 17 & 9207.6 & 9 & \0\0\04.6 &  &  &  &  \\
\\
	& 22 & 34061.9 & 0 & \0\015.5 &  &  &  &  \\
\\
$\qnfrac{3}{2}$ & 10 & 8539.6 & 0 &  & 1313.2 & 1248.7 & 1159.5 & 1046.5 \\
	& & & 1 &  & \0530.9 & \0475.1 & \0399.2 & \0305.7 \\
	& & & 2 &  & \0113.0 & \0\079.7 & \0\036.7 & \\
  \\
	& 11 & 8892.8 & 0 &  & \0842.7 & \0782.7 & \0700.0 & \0596.3 \\
	& & & 1 &  & \0347.7 & \0309.5 & \0257.6 & \0193.7 \\
	& & & 2 &  & \0\096.2 & \0\072.6 & \0\041.9 & \0\0\06.8 \\
	& & & 3 &  & \0\0\01.9 &  &  &  \\
	& 12 & 8892.8 & 6 &  & \0419.3 & \0355.3 & \0266.6 & \0154.3 \\
  \\
$\qnfrac{5}{2}$ & 4 & 4027.4 & 0 &  &  & \0\034.3 & \0\0\07.9 &  \\
	& 6 & 8892.8 & 17 &  &  & \0690.3 & \0617.4 & \0525.1 \\
	& & & 18 &  &  & \0169.2 & \0125.9 & \0\073.0 \\
	& & & 19 &  &  & \0\0\05.2 &  &  \\
\br
\end{tabular}
\end{indented}
\end{table}

\begin{table}
\caption{\label{tab:MC} Multichannel resonances, in units of MHz, of long-range 
	states with $P_T=+1$ in the
\mbox{${}^{3}$He(2 $^{3}$S$_{1}$)+${}^{4}$He(2 $^{3}$P$_{j}$)} and
${}^{3}$He(2 ${}^{3}$P$_{j}$) + ${}^{4}$He(2 ${}^{3}$S$_{1}$) systems which 
correspond to bound states obtained in the single-channel calculation.  Energies 
given are relative to the zero defined in table 1 and the next highest 
asymptotic energy is included in column $E_{n}^{\infty}$.
The column ``SC Contrib" indicates the index of the most strongly contributing 
adiabatic channel to that resonance (see text) and was always found to belong to 
the $\phi=\qnfrac{1}{2}$ adiabatic set. A dash indicates no particular channel had a 
dominant contribution. The column $E^\mathrm{match}_\mathrm{adi}$ gives the binding energy of a 
suggested assignment between the multichannel and single-channel calculations, 
when such an assignment is possible. The last column shows the difference in 
energy $\Delta E = E_\mathrm{MC} - E^\mathrm{match}_\mathrm{adi}$ between the resonance and 
single-channel level.} \begin{indented}
\lineup
\item[]\begin{tabular}{lllllll}
\br
$T$ & $E_{n}^{\infty}$ & $E_\mathrm{MC}$ & $\Gamma_\mathrm{MC}$ & SC Contrib & $E^\mathrm{match}_\mathrm{adi}$ & 
$\Delta E$ \\
\mr
$\qnfrac{1}{2}$ & 4027.4 & 3268.7 & 24.6 & 6 & -- & -- \\
	&        & 3791.9 & \07.3 & 7 & -- & -- \\
	&        & 3986.1 & 14.4 & 7 & -- & -- \\
	&        & 3997.4 & 0.29 & 7,8 & -- & -- \\
	& 8539.6 & 7976.5 & 31.6 & 13 & 517.0 & \0-45.5 \\
	&        & 8248.9 & 42.3 & 15 & 516.6 & -127.8 \\
	&        & 8476.3 & 12.8 & 19 & -- & -- \\
	&        & 8523.4 & 13.6 & 13 & \0\03.3 & \0-13.1 \\
	& 8892.8 & 8638.7 & 28.3 & 15 & 202.2 & \0-51.4 \\
	&        & 8651.7 & 27.7 & 15 & 202.2 & \0-38.4 \\
	&        & 8665.0 & 33.2 & 15 & 202.2 & \0-25.1 \\
	&        & 8683.6 & 38.4 & 15 & 202.2 & \0\0-6.5 \\
	&        & 8804.7 & 12.5 & 15 & \051.2  & \0-36.6 \\
	&        & 8872.6 & 25.4 & 16 & 133.1 & \0113.9 \\
	&        & 8892.0 & \02.4 & 15 & \051.2 & \0\050.5 \\
$\qnfrac{3}{2}$ & 4027.4 & 2387.2 & \06.6 & 4 & -- & -- \\
	&        & 3998.2 & 19.1 & 7 &  -- & -- \\
	&        & 4013.0 & \06.6 & 8 & -- & -- \\
	& 8539.6 & 8475.5 & 13.6 & 14 & 297.7 & \0232.6 \\
	&        & 8523.5 & 26.5 & 13 & 111.4 & \0\095.9 \\
	& 8892.8 & 8801.9 & 19.1 & 15 & 185.4 & \0\095.0 \\
	&        & 8861.7 & 22.2 & -- & -- & -- \\
	&        & 8888.9 & \04.2 & 15 & \039.8 & \0\036.2 \\
$\qnfrac{5}{2}$ & 4027.4 & 2476.3 & 10.5 & 4 & -- & -- \\
	&		 & 3860.4 & 17.6 & 7 & -- & -- \\
	&		 & 4006.9 & 14.7 & 8 & -- & -- \\
	& 8539.6 & 8152.6 & 20.3 & 14 & 252.7 & -135.1 \\
	&		 & 8531.3 & 15.6 & 13 & \080.4 & \0\072.5 \\
	& 8892.8 & 8715.4 & 16.2 & 15 & 158.2 & \0-18.5 \\
	&		 & 8861.0 & 13.3 & 15 & \022.1 & \0\0-9.8 \\
	&		 & 8883.8 & \06.1 & 15 & \022.1 & \0\013.0 \\
	&		 & 8888.4 & \05.4 & 15 & \022.1 & \0\017.6 \\
\br
\end{tabular}
\end{indented}
\end{table}

\begin{table}
\caption{\label{tab:MC_PTneg1} As for table \ref{tab:MC} but for resonances with 
$P_T=-1$.}
\begin{indented}
\lineup
\item[]\begin{tabular}{lllllll}
\br
$T$ & $E_{a}^{\infty}$ & $E_\mathrm{MC}$ & $\Gamma_\mathrm{MC}$ & SC Contrib &
$E^\mathrm{match}_\mathrm{adi}$ & $\Delta E$ \\
\mr
$\qnfrac{1}{2}$ & 4027.4 & 3766.6 & 40.2 & 12 & -- & -- \\
	&        & 3994.2 & \04.5 & -- & -- & -- \\
	&        & 4007.8 & 12.8 & 7,12 & -- & -- \\
	& 8539.6 & 6893.4 & 33.3 & 18 & -- & -- \\
	&        & 7951.2 & 42.2 & 19 & -- & -- \\
	&        & 8235.8 & 45.5 & 18 & -- & -- \\
	&        & 8369.3 & 49.0 & 18 & -- & -- \\
	&        & 8453.6 & 16.6 & 19 & -- & -- \\
	&        & 8532.8 & \07.3 & 13 & \0\03.3 & \0\0-3.7 \\
	& 8892.8 & 8788.3 & 16.0 & 19 & -- & -- \\
	&        & 8869.8 & 21.5 & 16 & 133.1 & \0111.4 \\
	&        & 8886.2 & \03.6 & 15 & \051.2 & \0\044.9 \\
$\qnfrac{3}{2}$ & 4027.4 & 3361.1 & 13.0 & -- & -- & -- \\
    &        & 4011.9 & \02.4 & 12,7 & -- & -- \\
	& 8539.6 & 8049.1 & 24.4 & 19 & -- & -- \\
	&        & 8517.2 & \08.4 & 19 & -- & -- \\
	& 8892.8 & 8830.8 & 15.5 & 19 & -- & -- \\
$\qnfrac{5}{2}$ & 4027.4 & 3727.7 & 24.6 & 20 & -- & -- \\
	&		 & 3996.0 & \00.2 & 8 & -- & -- \\
	&		 & 4019.7 & \08.5 & 7 & -- & -- \\
	& 8539.6 & 8466.5 & 23.1 & 18 & -- & -- \\
	&		 & 8513.4 & 10.4 & -- & -- & -- \\
	& 8892.8 & 8715.6 & 16.6 & 15 & 158.2 & \0-18.3 \\
	&		 & 8818.0 & 23.3 & 19 & -- & -- \\
	&		 & 8883.5 & \06.5 & -- & -- & -- \\
	&		 & 8887.6 & \04.8 & 15 & \022.1 & \0\016.8 \\
\br
\end{tabular}
\end{indented}
\end{table}
\subsection{Discussion}
The binding energies of the long-range states obtained using the single-channel 
calculation are listed in table \ref{tab:adi}.  Our multichannel calculations 
show a large collection of resonances which appear near the asymptotic energies 
of the Hamiltonian. Many of these have clearly originated from bound states that 
lie in the short-range wells of the single-channel potentials and are not of 
interest in this paper. There are other resonances, which are dominated by a 
long-range wave function and these have been listed in tables \ref{tab:MC} and 
\ref{tab:MC_PTneg1}. Although we have calculated the binding energies for bound 
levels in the adiabatic potentials for up to $T=\qnfrac{9}{2}$, we have only gone as far 
as $T=\qnfrac{5}{2}$ in the multichannel calculations as higher values of $T$ introduce no 
significant changes in the structure of the couplings.

All of the bound levels and resonances found can be placed into three groups, 
depending on which asymptote they are closest to. One of these groups consists 
of
two isolated levels only, of energy approximately $34047.8$~MHz and $9202.9$~MHz, 
which were found in the single-channel but not in the multichannel calculations.  
The rest of the bound levels and resonances are either close to the 
$4027.4$~MHz, $8539.6$~MHz or $8892.8$~MHz asymptotes.

\subsubsection{Group 1: Resonances beneath the $4027.4$~MHz asymptote}

This asymptote corresponds to the bosonic ${}^4$He in its $2s$ state and the 
fermionic ${}^3$He in the $2p$ state with $f=\qnfrac{3}{2}$ and $\tilde{j} = 1$.  
In the single-channel calculation, a long-range well was found in the set of 
adiabatic potentials. However, after calculating the bound states supported by 
this well, we observed a strong tunneling out of the long-range well into the 
short-range region and so rejected these states as long-range candidates.  In 
the multichannel calculation, however, we find relatively long-range resonances 
whose widths are not too large. This leads us to conclude that the adiabatic 
potentials must be significantly coupled by non-adiabatic terms such that the 
adiabatic avoided crossings between the potentials become, in the multichannel 
calculation, true crossings.

By revisiting these adiabatic potentials, and forcing a few non-adiabatic 
crossings between different potentials, we have found two additional bound 
levels at 3765.30~MHz and 3999.71~MHz (with binding energies of 262.07~MHz and 
27.66~MHz respectively), which correspond to a pair of the multichannel 
resonances.  However, there still remain many more resonances than can be 
observed in the single-channel calculations. Many of these resonances also have 
relatively small line widths.

\subsubsection{Group 2: Resonances beneath the $8539.6$~MHz and $8892.8$~MHz 
asymptotes}

For these two asymptotes, the $2p$ excitation can be found on either atom -- 
the $8539.6$~MHz asymptote corresponds to a $2s$~${}^4$He atom and a 
$2p$~${}^3$He atom with $f=\qnfrac{1}{2}$ and $\tilde{j}=1$, and the asymptote 
$8892.8$~MHz corresponds to a $2s$~${}^3$He with $f=\qnfrac{1}{2}$ and a 
$2p$~${}^4$He with $j=2$.

To make comparisons between the single-channel and multichannel results, we have
transformed the multichannel resonance wave functions into the adiabatic basis 
(\ref{cpw35}) and performed the integrals $\int_0^{R_\mathrm{max}} |G_n(R)|^2 
dR$, where $R^{-1} G_n(R)$ is the transformed wave function, in order to obtain 
the contribution that each adiabatic channel makes to that resonance.  However, 
as the wave functions increase exponentially for large $R$ and are not $L^2$ 
normalizable, this method alone is not well defined.  To obtain a useful measure 
of the adiabatic contributions, we fit the asymptotic shape of the function 
$|G_n(R)|^2$ to a form $f^\mathrm{fit}_n(R) = A \exp(-\mathrm{Im}(k_n)R)$ where 
$\hbar k_n = \sqrt{2\mu(\mathcal{E} - E^\infty_n)}$\footnote{Note that 
$-\mathrm{Im}(k_n) > 0$ always.} and subtract an amount $\int_0^{R_\mathrm{max}} 
f^\mathrm{fit}_n(R) dR$ from these contributions.  Although this process 
destroys the positivity of the normalization, it provides a sufficiently clear
set of relative contributions from each of the adiabatic channels. This allows
us to identify the particular adiabatic channel (if it exists) that is mainly
responsible for each multichannel resonance.

With these adiabatic contributions, we can make some assignments between the 
single-channel and multichannel results. The first note-worthy feature is that 
none of the $\phi = \qnfrac{3}{2}$ or $\phi = \qnfrac{5}{2}$ single-channel levels seem to have 
survived the couplings to open channels. Although these adiabatic channels do 
contribute, they are never dominant.

In contrast, a couple of the adiabatic potentials ($k=15,16$) in the $\phi=\qnfrac{1}{2}$ 
set can almost entirely be identified with multichannel resonances. Although 
visually there is little to distinguish these adiabatic potentials from the 
others, we note that their minima occur at relatively small distances 
($R\approx259\,a_0$ and $R\approx276\,a_0$) as opposed to the minima for the 
other potentials ($R=300\,a_0$ to $400\,a_0$). We also observe two clusters of 
resonances that appear to correspond to one single-channel level in both the 
$T=\qnfrac{1}{2}$ and $T=\qnfrac{5}{2}$ sets at binding energies of $202.2$~MHz and $22.0$~MHz 
respectively. It is not known why these clusters have appeared but they can 
represent an interesting regime for experiment to probe, although they suffer 
from relatively large line widths.

We note that the energies differ significantly between the single-channel and 
multichannel results and it is only with the assistance of the adiabatic 
contributions that we were able to make these assignments. The difference in 
energy $\Delta E = E_\mathrm{MC} - E_\mathrm{adi}$ is included in the last 
column of tables \ref{tab:MC} and \ref{tab:MC_PTneg1}.

Although there is often a similarity between the $P_T=+1$ and $P_T=-1$ results, 
it is clear that $P_T=-1$ has far fewer assignments to the single-channel 
results. In fact, the $\phi=\qnfrac{1}{2}$ adiabatic potentials $k=18,19$ seem to be the 
origin for many of these resonances in the $P_T=-1$ set and yet these adiabatic 
potentials do not support bound levels when non-adiabatic and Coriolis couplings 
are ignored.

\subsubsection{Short-range resonances}

We conclude this section with a comment about the multitude of other levels that
were obtained in the multichannel calculations, but which have not been reported
in this paper. These levels are easily identifiable, as they appear in a closely
packed sequence of energies whose resonance wave function probabilities are 
reasonably uniformly distributed to all radial distances. Any visible peaks in 
their probabilities are concentrated at small radial distances and are usually 
spread over a large range of adiabatic channels.

There are two reasons why we do not discuss these resonances further.  Firstly, 
the uncertainty in the input potentials is at its largest for short ranges and 
secondly, these levels will be altered by the process of Penning ionization that 
is likely to dramatically decrease the lifetime of these resonances when they 
are non-negligible for $R \lesssim 7~a_0$.

Although all of the long-range multi-channel levels that we identified exist 
only near the $E^\infty_a = 4027.4$~MHz, $E^\infty_a = 8539.6$~MHz and 
$E^\infty_a = 8892.8$~MHz asymptotes, we wish to emphasize that we have scanned 
all asymptotes in the multichannel calculations and found no additional 
long-range resonances.

\section{Conclusions}
Long-range bound states of the excited heteronuclear 
${}^{3}$He$^{*}$--${}^{4}$He$^{*}$ system that dissociate to either 
${}^{3}$He(1s2s~${}^{3}$S$_{1}$) + ${}^{4}$He(1s2p~${}^{3}$P$_{j}$) or
${}^{3}$He(1s2p~${}^{3}$P$_{j}$) + ${}^{4}$He(1s2s~${}^{3}$S$_{1}$),
where \mbox{$j=0, 1, 2$}, have been investigated using both single-channel and multichannel 
calculations in order to analyse the effects of Coriolis and non-adiabatic 
couplings.

In the single-channel calculation, several long-range wells were found in the 
sets of adiabatic potentials which supported a large number of bound states.  In 
addition, the full set of coupled equations of the multichannel problem were 
solved by extending the calculations to allow for the identification of 
resonances with finite lifetimes.  A large number of resonances near to the 
asymptotes of 4027.4~MHz, 8539.6~MHz and 8892.8~MHz have been predicted and many 
of these could be identified with bound levels in the single-channel calculation 
in which non-adiabatic and Coriolis couplings were neglected.  

As all of these assignments of multichannel resonances with single-channel bound 
levels coincided with a quantum number of $\phi=|\Omega_f|=\qnfrac{1}{2}$, we are 
lead to conclude that it is important to consider the full multichannel set of 
equations in order to even qualitatively describe the spectroscopy of the 
system. This is in contrast to the homonuclear systems, in which a stronger link 
between single-channel and multichannel states was found. The largest 
discrepancy was observed for multichannel resonances of total parity $P_T = -1$, 
where only a rare few resonances could be identified with single-channel levels.

There were no purely bound states identified in the calculations as all of the 
resonances lie above the lowest asymptote of 2153.1~MHz. However, this is not 
surprising due to the large number of asymptotes available to the heteronuclear 
configuration, allowing many opportunities for predissociation. 
These resonances arise from bound states of one channel being embedded in the 
continuum of another channel and, in that sense, are Feshbach in nature.
Fortunately, some 
resonances do have a very small line width of less than 1~MHz near to the 
asymptote of 4027.4~MHz, which could prove to be very useful in 
photoassociation experiments.

In our previous publications that addressed the homonuclear systems of 
metastable helium collisions we were able to provide a set of observability 
criteria, which indicated a likelihood for colliding metastable atoms to be 
photoassociated into the resonance. Unfortunately there is not a clear and 
obvious choice for the preparation of a heteronuclear experimental gas mixture, 
so we have not included a set of observability criteria in this paper. It would 
be desirable, in such a case, to perform a scattering calculation similar to 
\cite{Cocks09}, which would consider the appropriate incoming $2s+2s$ channels 
and laser coupling terms. The full scattering matrix could then be obtained, 
along with various cross sections relevant to experiment.

We also note that a discussion of predissociation widths due to the presence of 
Penning ionization is essential to make predictions for experiment.  
Fortunately, Penning ionization only occurs at short ranges $\lesssim 7~a_0$ so 
its impact on the long-range resonances that we have identified should be small 
and likely negligible. 

\appendix

\section{Basis states and matrix elements}
\label{app:states}

The body-fixed (molecular) states in the coupling scheme (\ref{cpw11}) are
\begin{eqnarray}
\label{cpwa1}
\fl
|(\gamma_{1}j_{1}i_{A}f_{1A})_{A}, (\gamma_{2}j_{2}i_{B}f_{2B})_{B}, f, 
\Omega_{f},  T, m_{T} \rangle  \nonumber  \\
= |T,m_{T}, \Omega_{f} \rangle
\sum_{\Omega_{f_{1A}}\Omega_{f_{2B}}}\;
\sum_{\Omega_{j_{1}}\Omega_{j_{2}}} \;
\sum_{\Omega_{i_{1}}\Omega_{i_{2}}} \;
\sum_{\Omega_{L_{1}}\Omega_{L_{2}}}  \;
\sum_{\Omega_{S_{1}}\Omega_{S_{2}}}
\nonumber  \\  
\times
\Cleb{f_{1A}}{f_{2B}}{f}{\Omega_{f_{1A}}}{\Omega_{f_{2B}}}{\Omega_{f}}
\Cleb{j_{1}}{i_{1}}{f_{1A}}{\Omega_{j_{1}}}{\Omega_{i_{1}}}{\Omega_{f_{1A}}}
\Cleb{j_{2}}{i_{2}}{f_{2B}}{\Omega_{j_{2}}}{\Omega_{i_{2}}}{\Omega_{f_{2B}}}
\Cleb{L_{1}}{S_{1}}{j_{1}}{\Omega_{L_{1}}}{\Omega_{S_{1}}}{\Omega_{j_{1}}}
\nonumber  \\ 
 \times
\Cleb{L_{2}}{S_{2}}{j_{2}}{\Omega_{L_{2}}}{\Omega_{S_{2}}}{\Omega_{j_{2}}}
|\gamma_{1}\Omega_{L_{1}}\Omega_{S_{1}}\rangle_{A}
|i_{A}\Omega_{i_{A}} \rangle
\nonumber  \\ 
\times  |\gamma_{2}\Omega_{L_{2}}\Omega_{S_{2}}\rangle_{B}
|i_{B}\Omega_{i_{B}} \rangle
\end{eqnarray}
where the transformation between the molecular and space-fixed states is, 
for example,
\begin{equation}
\label{cpwa2}
|j \Omega_{j} \rangle = \sum_{m_{j}} 
D^{j}_{m_{j}\Omega_{j}}(\varphi, \theta ,0) |j m_{j} \rangle 
\end{equation}
and $\Cleb{j_1}{j_2}{j}{m_1}{m_2}{m}$ is a Clebsch-Gordan coefficient.

Introducing the coupled states
\begin{eqnarray}
\label{cpwa2a}
\fl
|\gamma_{1}\Omega_{L_{1}}\Omega_{S_{1}}\rangle_{A}
|\gamma_{2}\Omega_{L_{2}}\Omega_{S_{2}}\rangle_{B}
= \sum_{LS\Omega_{L}\Omega_{S}}
\Cleb{L_{1}}{L_{2}}{L}{\Omega_{L_{1}}}{\Omega_{L_{2}}}{\Omega_{L}}
\Cleb{S_{1}}{S_{2}}{S}{\Omega_{S_{1}}}{\Omega_{S_{2}}}{\Omega_{S}}
\nonumber  \\
\times |(\gamma_{1})_{A}, (\gamma_{2})_{B}, L S \Omega_{L} \Omega_{S} \rangle 
\end{eqnarray}
and
\begin{equation}
\label{cpwa3}
|i_A, i_B, i \Omega_{i} \rangle =\sum_{\Omega_{i_A}\Omega_{i_B}} 
\Cleb{i_A}{i_B}{i}{\Omega_{i_A}}{\Omega_{i_B}}{\Omega_{i}}
|i_A \Omega_{i_A} \rangle |i_B \Omega_{i_B} \rangle
\end{equation}
and expressing the sums over Clebsch-Gordan coefficients in terms of
Wigner 9-$j$ symbols  \mbox{\footnotesize{$\Ninej{a}{b}{c}{d}{e}{f}{g}{h}{i}$}} 
gives
\begin{eqnarray}
\label{cpwa4}
\fl
|(\alpha_{1})_{A}, (\alpha_2)_{B}, f, \Omega_{f},  T, m_{T} \rangle  \nonumber  
\\
= |T,m_{T}, \Omega_{f} \rangle 
\sum_{i \Omega_{i}} \sum_{j \Omega_{j}} \sum_{LS\Omega_{L}\Omega_{S}}
F_{ji\Omega_{j}\Omega_{i}}^{f_{1A}f_{2B}f\Omega_{f}}   
F_{LS\Omega_{L}\Omega_{S}}^{j_{1}j_{2}j\Omega_{j}}  \nonumber  \\ \times
|(\gamma_{1})_{A}, (\gamma_{2})_{B}, L S \Omega_{L} \Omega_{S} \rangle 
|i_A,i_B,i\Omega_{i}\rangle,
\end{eqnarray}
where $(\alpha_{i})_{X}\equiv \{\gamma_{i},j_{i},i_{X},f_{iX}\}$.
The coupling coefficients 
are defined by
\begin{equation}
\label{cpwa5}
\fl
F_{LS\Omega_{L}\Omega_{S}}^{j_{1}j_{2}j\Omega_{j}} =  
[LSj_{1}j_{2}]^{\frac{1}{2}}
\Cleb{L}{S}{j}{\Omega_L}{\Omega_S}{\Omega_j}
\Ninej{L_1}{L_2}{L}{S_1}{S_2}{S}{j_1}{j_2}{j} 
\end{equation}
and
\begin{equation}
\label{cpwa5a}
\fl
F_{ji\Omega_{j}\Omega_{i}}^{f_{1A}f_{2B}f\Omega_{f}} =  
[jif_{1A}f_{2B}]^{\frac{1}{2}}
\Cleb{j}{i}{f}{\Omega_j}{\Omega_i}{\Omega_f}
\Ninej{j_1}{j_2}{j}{i_A}{i_B}{i}{f_{1A}}{f_{2B}}{f} \end{equation}
where $[ab\ldots ]=(2a+1)(2b+1)\ldots $.
In (\ref{cpwa5}) and (\ref{cpwa5a}) the implicit set of quantum numbers 
$(\gamma_{1},\gamma_{2})$ and $(j_{1},i_{A},j_{2},i_{B})$ respectively have been 
suppressed. 

The eigenstates of total parity $\hat{P}_{T}$ \cite{CPW11} are:
\begin{eqnarray}
\label{cpwa51}
\fl
|(\alpha_{1})_{A}, (\alpha_{2})_{B}, f, \phi ,T, m_{T}; P_{T} \rangle
= N_{P_{T}} [|(\alpha_{1})_{A}, (\alpha_{2})_{B}, f, \phi ,T, m_{T}\rangle
\nonumber \\
+P_{T}P_{1}P_{2}(-1)^{f-T}
|(\alpha_{1})_{A}, (\alpha_{2})_{B}, f, -\phi ,T, m_{T}\rangle ] ,
\end{eqnarray}
where $P_{i}=(-1)^{L_{i}}$ is the parity of the atomic state 
$|L_{i}m_{L_{i}}\rangle $, $\phi \equiv |\Omega_{f}| = |\Omega_{T}|$ and 
$N_{P_{T}}=1/\sqrt{2}$. The relationship to the $LS$ basis is completed by using 
(\ref{cpwa4}).

Defining
\begin{eqnarray}
\label{cpwa6}
\tilde{F}^{12}_{AB}
&=\sum_{\Omega_{j}}F_{ji\Omega_{j}\Omega_{i}}^{f_{1A}f_{2B}f\Omega_f}
F_{LS\Omega_{L}\Omega_{S}}^{j_{1}j_{2}j\Omega_{j}} \nonumber \\
&= \sum_{\Omega_j} [LSj_1 j_2 ji f_{1A} f_{2B}]^{1/2} \nonumber \\
&\times
\Cleb{j}{i}{f}{\Omega_j}{\Omega_i}{\Omega_f} 
\Cleb{L}{S}{j}{\Omega_l}{\Omega_S}{\Omega_j} 
\Ninej{j_1}{j_2}{j}{i_A}{i_B}{i}{f_{1A}}{f_{2B}}{f}
\Ninej{L_1}{L_2}{L}{S_1}{S_2}{S}{j_1}{j_2}{j}
\end{eqnarray}
then gives (\ref{cpw19}). The state with $1 \leftrightarrow 2$ is obtained by 
reordering the angular momenta to give
\begin{eqnarray}
\label{cpwa61}
\fl
|(\alpha_{2})_{A}, (\alpha_{1})_{B}, f, \Omega_f ,T, m_{T} \rangle
\nonumber  \\
= |T, m_{T}, \Omega_f \rangle \sum_{i \Omega_{i} j} 
\sum_{LS\Omega_{L}\Omega_{S}} \tilde{F}^{21}_{AB}
\nonumber  \\
\times |(\gamma_{2})_{A}, (\gamma_{1})_{B}, L S \Omega_{L} \Omega_{S} \rangle 
|i_{A}, i_{B}, i \Omega_{i} \rangle.
\end{eqnarray}

The matrix element of $\hat{H}_{\mathrm{el}}$ in the basis (\ref{cpwa4})
is then
\begin{eqnarray}
\label{cpwa7}
\fl
{}_{12}\langle \tilde{a}^{\prime}, 
\Omega_f^\prime{=}\phi^{\prime}|\hat{H}_{\mathrm{el}}
|\tilde{a} ,\Omega_f{=}\phi \rangle_{12}  = \delta_{\xi, \xi^{\prime}}
\sum_{i^{\prime} \Omega_{i}^{\prime} j^{\prime}} 
\sum_{L^{\prime}S^{\prime}\Omega_{L}^{\prime}\Omega_{S}^{\prime}}
\sum_{i \Omega_{i} j} \sum_{LS\Omega_{L}\Omega_{S}}
\tilde{F}^{1^\prime 2^\prime}_{AB}
\tilde{F}^{12}_{AB} \nonumber \\
\times \langle (\gamma_{1}^{\prime})_{A}, (\gamma_{2}^{\prime})_{B}, 
L^{\prime} S^{\prime} \Omega_{L}^{\prime} \Omega_{S}^{\prime}  
|\hat{H}_{\mathrm{el}}
|(\gamma_{1})_{A}, (\gamma_{2})_{B}, L S \Omega_{L} \Omega_{S} \rangle 
\end{eqnarray}
where $\tilde{a} \equiv \{(\alpha_{1})_{A}, (\alpha_{2})_{B},f,T,m_{T}\}$ and
$\xi \equiv \{i_{A},i_{B},\phi, T, m_{T} \}$.  The results for the matrix 
elements ${}_{21}\langle \tilde{a}^{\prime}, \phi^{\prime}|\hat{H}_{\mathrm{el}}
|\tilde{a}, \phi \rangle_{21}$ and
${}_{21}\langle \tilde{a}^{\prime}, \phi^{\prime}|\hat{H}_{\mathrm{el}}
|\tilde{a}, \phi \rangle_{12}$ can be obtained from (\ref{cpwa7}) by appropriate 
substitution of the $(1,2)$ labels.  The matrix elements for the $(\phi, 
\phi^{\prime}) \rightarrow (-\phi, -\phi^{\prime})$ cases can be shown to be
\begin{equation}
	\label{cpwa7b}
	\fl
	{}_{12}\langle \tilde{a}^\prime,\Omega_f^\prime{=}-\phi^\prime | 
	H_\mathrm{el} | \tilde{a}, \Omega_f{=}-\phi\rangle_{12} = (-1)^{f-f^\prime} 
	{}_{12}\langle \tilde{a}^\prime,\Omega_f^\prime{=}\phi^\prime | 
	H_\mathrm{el} | \tilde{a}, \Omega_f{=}\phi\rangle_{12},
\end{equation}
and, due to the Kronecker delta in $\phi$,
\begin{equation}
	\label{cpwa7c}
	{}_{12}\langle \tilde{a}^\prime,\Omega_f{=}\phi^\prime| H_\mathrm{el} 
	|\tilde{a},\Omega_f{=}-\phi \rangle_{12} = 0.
\end{equation}
The same relationship is found for the $21$-$12$ and $21$-$21$ cases.
The phase factor $(-1)^{f-f^\prime}$ conveniently cancels out in the basis 
(\ref{cpwa51}), due to the presence of a Kronecker delta in $T$, leaving the 
result:
\begin{eqnarray}
	\fl
	{}_{12}\langle \tilde{a}^\prime,\phi^\prime,P_T^\prime|H_\mathrm{el} 
	|\tilde{a},\phi,P_T\rangle_{12} \nonumber \\
	= \frac{1}{2} \left[ {}_{12}\langle \tilde{a},\Omega_f^\prime{=}\phi^\prime| 
		H_\mathrm{el} | \tilde{a},\Omega_f{=}\phi\rangle_{12} \right. \nonumber 
		\\
		\left. + (-1)^{f-f^\prime}(-1)^{T-T^\prime} P_T P_T^\prime 
	{}_{12}\langle\tilde{a}^\prime,\Omega_f^\prime{=}-\phi^\prime | 
H_\mathrm{el} |\tilde{a},\Omega_f{=}-\phi\rangle_{12} \right] \nonumber \\
	= \delta_{P_T,P_T^\prime}\, {}_{12}\langle 
	\tilde{a}^\prime,\Omega_f^\prime{=}\phi^\prime| H_\mathrm{el} 
	|\tilde{a},\Omega_f{=}\phi\rangle_{12}.
\end{eqnarray}
The same is true also for $1\leftrightarrow 2$.  The necessary matrix elements of $\hat{H}_{\mathrm{el}}$ can be expressed in 
terms of the BO potentials for the homonuclear systems using
\begin{equation}
\label{cpwa8}
\fl
|LS\rangle_{12} \equiv |(\gamma_{1})_{A}, (\gamma_{2})_{B}, L S \Omega_{L} 
\Omega_{S} \rangle =N_{w} [|g \rangle + |u \rangle ]
\end{equation}
and
\begin{equation}
\label{cpwa9}
\fl
|LS\rangle_{21} \equiv |(\gamma_{2})_{A}, (\gamma_{1})_{B}, L S \Omega_{L} 
\Omega_{S} \rangle =N_{w} \varepsilon_{LS}[|g \rangle - |u \rangle ]
\end{equation}
where we have introduced the notation
\begin{equation}
\label{cpwa10}
\fl
|g \rangle = |\gamma_{1}\gamma_{2}, L S \Omega_{L} \Omega_{S};g \rangle , 
\qquad
|u \rangle = |\gamma_{1}\gamma_{2}, L S \Omega_{L} \Omega_{S};u \rangle 
\end{equation}
for the homonuclear eigenstates of \textit{gerade} and \textit{ungerade} 
symmetry. This gives
\begin{eqnarray}
\label{cpwa11}
\fl
{}_{12} \langle L^{\prime} S^{\prime} |\hat{H}_{\mathrm{el}}
| L S \rangle _{12} =
{}_{21} \langle L^{\prime} S^{\prime} |\hat{H}_{\mathrm{el}}
| L S \rangle _{21}   \nonumber  \\
= \delta_{\nu, \nu^{\prime}} \frac{1}{2} 
[{}^{2S+1}\Lambda_{g}^{+}(R)+{}^{2S+1}\Lambda_{u}^{+}(R)+
2E_{\Lambda S}^{\infty}]
\end{eqnarray}
and
\begin{equation}
\label{cpwa12}
\fl
{}_{21} \langle L^{\prime} S^{\prime} |\hat{H}_{\mathrm{el}}
| L S \rangle _{12}  = \delta_{\nu, \nu^{\prime}} \varepsilon_{LS} \frac{1}{2} 
[{}^{2S+1}\Lambda_{g}^{+}(R)-{}^{2S+1}\Lambda_{u}^{+}(R)]
\end{equation}
where $\nu \equiv \{\gamma_{1}, \gamma_{2}, L, S, \Omega_{L}, \Omega_{S} \}$.

\section{Determination of Resonances}
\label{app:MC}

To determine the position of resonances, which have the form $\mathcal{E} = E - 
i\Gamma$, we choose to scan the region of complex energy space around each 
asymptote with a contour integral approach. For each point $\mathcal{E}$ that is 
visited, we perform an inward and an outward integration of the multichannel 
equations (\ref{cpw7}), starting from $R=R_\mathrm{max}$ and $R=R_\mathrm{min}$ 
respectively and setting closed (open) boundary conditions for the channels 
below (above) the real component of $\mathcal{E}$ \footnote{Note that this 
requires that no asymptote energy lies inside a contour, so that the boundary 
conditions do not change and the contour integrand remains analytic.}.  These 
integrations end at a common point $R_\mathrm{mid}$ which allows for the 
definition of a matching condition $D(\mathcal{E})=0$ \cite{CWP10}, which is 
satisfied only at the location of a resonance.  The integrand of the contour 
integral is chosen to be $f(\mathcal{E})=1/D(\mathcal{E})$, such that the poles 
of $f(\mathcal{E})$ designate the positions of the resonances.  Numerical tests 
have shown that $f(\mathcal{E})$ is analytic away from resonances and the 
asymptotes, which we believe is due to a nontrivial relationship to the 
propagator $(\hat{H} - \mathcal{E})^{-1}$.

Due to the analyticity of $f(\mathcal{E})$, we can use Cauchy's residue theorem
\begin{equation}
	\oint_C f(\mathcal{E}) d\mathcal{E} = 2\pi i \sum \mathrm{residues}.
\end{equation}
This enables a clear identification of the presence of a residue and 
consequently a resonance within a region of $\mathcal{E}$-space
\footnote{We note that there is the mathematical possibility for a contour to 
contain two resonances with residues which additively cancel, however our tests 
have shown that each resonance has a distinctly different residue.}.  Using
this, we may very quickly narrow the search to individual regions which tightly
bound a single resonance. 

For each asymptote $a$ with energy $E^\infty_a$, we start with a contour 
integration over a large box in complex $\mathcal{E}$ space, with the real part 
spanning $E^\infty_a - \delta$ to $E^\infty_a - 2000$~MHz, where $\delta$ is a 
small parameter (we choose $\delta=0.5$~MHz) that avoids the non-analytic 
behaviour of the change in boundary conditions at the asymptote. The imaginary 
part of the box is chosen to span the range $+\delta$ to $-50$~MHz, where 
$\delta$ is included so that true bound states do not intersect the edge of the 
contour.  By subdividing this box, progressively narrowing the span of the real 
and imaginary parts, we can eliminate regions of $\mathcal{E}$-space that 
contain no resonances and continue until each resonance has been identified to 
an accuracy of $1$~MHz, after which we switch to a gradient descent method to 
obtain the final accuracy desired.

For each box contour integration there are four separate line integrals 
$L_{i}, i=1,2,3,4$ which we perform adaptively via a Gaussian quadrature method.  
The final accuracy required, however, is that of the sum $S = \sum_i L_i$ of all 
four integrals and this can be of the same order as the integration error 
itself.  This is especially true for contours which contain no poles. To obtain 
the desired relative error in $S$, denoted by $\epsilon_S$, we perform the 
integration iteratively. We first obtain an estimate of the value of $S\approx 
S^{(0)}$ and then, using the computed desired absolute error $\delta S^{(0)} = 
S^{(0)} \epsilon_S$, we update the desired relative error for each of the line 
integrals to be $\epsilon^{(1)}_{L_i} = \delta S^{(0)} / 4 L_i$.  Performing the 
line integrals again returns a new value $S^{(1)}$ which closes the iteration 
loop.\footnote{Note that we cache function values to avoid unnecessary 
repetition of calculations.} There are two stopping conditions for this loop: 
(i) $\delta S / S < \epsilon_S$ which corresponds to a non-zero residue, and 
(ii) we reach machine precision while specifying an updated tolerance 
$\epsilon^{(j)}_{L_i}$.  The later condition corresponds to either a non-zero 
residue that is much smaller than the integrand (in which case we are forced to 
neglect it) or it corresponds to our best representation of an integration of 
zero. To this end, we set a small tolerance ($\epsilon_\mathrm{machine} = 
10^{-8}$) which is taken relative to the total value of all line integrals 
$\sum_i |L_i|$ and any result $S^{(i)} < \epsilon_\mathrm{machine} \sum_i |L_i|$ 
is assumed to be zero.

There is an alternative method to identify the resonances, which was used in our 
previous publications \cite{CWP10,CPW11}, namely Cauchy's argument principle 
which replaces the integrand of the contour integration by its logarithmic 
derivative. The integral is then equal to the difference in the number of zeros  
and poles, that is
\begin{equation}
	\frac{1}{2\pi i} \oint_C \frac{f^\prime(\mathcal{E})}{f(\mathcal{E})} 
	d\mathcal{E} = n_\mathrm{zeros} - n_\mathrm{poles}.
\end{equation}
The advantage of this method is that the integration results in integer values, 
allowing a clear distinction between contours with and without a resonance.  
There is also no possible issue of two resonances with equal and opposite 
residues cancelling out.  However, the disadvantage is that the method can 
``hide" resonances when an equal number of zeros and poles lie in one region.  
By subdividing the region into a fine grid of box contours, we were confident 
that all resonances had been identified. However, with the increased number of 
channels in the current ${}^{3}$He$^{*}$--${}^{4}$He$^{*}$ system, this becomes 
prohibitively expensive due to a far greater number of pairs of poles and zeros.

\end{document}